# Joule-Thomson cooling of $CO_2$ injected into aquifer under heat exchange with adjacent formations by Newton's law- 1D exact solution.

Christina Chesnokov[1], Rouhi Farajzadeh[2,3], Kofi Ohemeng Kyei Prempeh[1], Siavash Kahrobaei[2], Jeroen Snippe[2], Pavel Bedrikovetsky[1]

[1] The University of Adelaide, Australia

[2] Shell Global Solutions International, The Hague, The Netherlands

[3] Delft University of Technology, The Netherlands

**Abstract**

This paper discusses axi-symmetric flow during $CO_2$ injection into a non-adiabatic reservoir accounting for Joule-Thomson cooling and steady-state heat exchange between the reservoir and the adjacent layers by Newton's law. An exact solution for this 1D problem is derived and a new method for model validation by comparison with quasi 2D analytical heat-conductivity solution is developed. The temperature profile obtained by the analytical solution shows a temperature decrease to a minimum value, followed by a sharp increase to initial reservoir temperature. The analytical model exhibits stabilisation of the temperature profile and cooled zone. When accounting for heat gain from adjacent layers, the minimum temperature is higher compared to the case with no heat exchange. The analytical model can be used to evaluate the risks of hydrate formation and/or rock integrity during $CO_2$ storage in depleted reservoirs.

**Nomenclature**

| | | |
|---|---|---|
| $a_a$ | Thermal diffusivity of adjacent formations | m²/s |
| $A^*$ | Dimensionless J-T number | [-] |
| $B$ | Dimensional heat exchange | m⁻³ |
| $C$ | Dimensional heat exchange for steady state case | m⁻² |
| $D$ | Dimensional J-T number for steady state case | K m |
| $c_s$ | Heat capacity of reservoir rock | J/kg/K |
| $c_w$ | Heat capacity of water | J/kg/K |

| Symbol | Description | Units |
|---|---|---|
| $c_f$ | Heat capacity of $CO_2$ | J/kg/K |
| $c_{ra}$ | Heat capacity of adjacent formation rock | J/kg/K |
| $c^{res}$ | Overall reservoir heat capacity | J/m³/K |
| $c^a$ | Overall adjacent formation heat capacity | J/m³/K |
| $H$ | Formation thickness | m |
| $k$ | Permeability | m² |
| $k_{rf}^e$ | End-point relative permeability of $CO_2$ | [-] |
| $l$ | Shale thickness | m |
| $M_J$ | Mass injection rate | kg/s |
| $p$ | Fluid pressure | Pa |
| $p_J$ | Injection pressure of $CO_2$ | Pa |
| $p_w$ | Reservoir pressure | Pa |
| $q_J$ | $CO_2$ injection rate | m³/s |
| $r$ | Radial distance along flow direction | m |
| $r_D$ | Dimensionless radial distance along flow direction | [-] |
| $r_w$ | Well radius | m |
| $r_f$ | $CO_2$ front position | m |
| $r_{norm}$ | Normalized distance | [-] |
| $r_{inj}$ | Injection radius | m |
| $r_p$ | Temperature penetration depth radius | m |
| $r_c$ | Critical radius of validity | [-] |
| $S_{wi}$ | Connate water saturation | [-] |
| $T$ | Temperature | K |
| $T_I$ | Initial reservoir temperature | K |
| $T_J$ | Injection temperature | K |
| $T_D$ | Dimensionless reservoir temperature | [-] |
| $T_{min}$ | Minimum reservoir temperature | K |
| $t$ | Time | s |
| $t_0$ | Starting time for heat propagation | s |
| $t_{inj}$ | Injection time | s |
| $t_D$ | Dimensionless time | [-] |
| $t_{D\,max}$ | Maximum dimensionless time for validity | [-] |
| $V$ | Dimensional heat front velocity | m$^{-1}$ |
| $V^*$ | Dimensionless heat front velocity | [-] |
| $x$ | Spatial integral variable | m |
| $z$ | Distance along vertical direction for heat flow between layers | m |
| $z_D$ | Dimensionless vertical coordinate | [-] |
| **Greek letters** | | |
| $\alpha_{JT}$ | Joule-Thomson coefficient | K/Pa |
| $\gamma_{ra}$ | Heat conductivity of adjacent layers | W/m/K |
| $\gamma_{rs}$ | Heat conductivity of shale | W/m/K |
| $\gamma_w$ | Heat conductivity of water | W/m/K |
| $\gamma^s$ | Overall shale heat conductivity | W/m/K |
| $\gamma^a$ | Overall adjacent layer heat conductivity | W/m/K |
| $\theta$ | Dimensionless temperature in vertical heat conduction | [-] |
| $\kappa$ | Overall heat transfer coefficient of shale | W/m$^2$/K |
| $\mu_f$ | Viscosity of $CO_2$ | Pa s |
| $\rho_s$ | Reservoir Rock density | kg/m$^3$ |
| $\rho_w$ | Water density | kg/m$^3$ |
| $\rho_f$ | $CO_2$ density | kg/m$^3$ |
| $\rho_{ra}$ | Adjacent layers rock density | kg/m$^3$ |
| $\tau$ | Temporal integral variable | s |
| $\phi$ | Reservoir Porosity | [-] |
| $\phi_s$ | Shale Porosity | [-] |
| $\phi_a$ | Adjacent layers Porosity | [-] |
| $\omega$ | Dimensionless Newton's validity parameter | [-] |

1. **INTRODUCTION**

The growing concerns over negative impacts of greenhouse gases, in particular Carbon dioxide ($CO_2$), on climate change have increased the interest in developing different methods to reduce their emissions into the atmosphere. A practical and complementary solution is to capture $CO_2$ from highly concentrated sources, e.g., industrial plants or directly from air and store it in underground geological formations such as depleted oil and gas reservoirs, deep saline aquifers, and coal beds.

Depleted gas fields exhibit several advantages over other geological formations. The extracted gas from these fields can in principle be replaced by $CO_2$ and as such, huge volumes of $CO_2$ can be stored in these reservoirs. For example, it has been estimated that in the Dutch sector of the North Sea, theoretically, more than 1.5 Gton of $CO_2$ can be stored in the depleted gas fields (Van der Velde et al., 2008). Due to large recovery factors (extracted fraction of the initial volumes) and compressibility of gas more space is available in the gas fields compared to oil fields, in which the extracted oil is usually replaced by injection of another (usually incompressible) fluids (Hamza et al., 2021).

The production of oil fields often involves injection of external fluids (for pressure maintenance or improved-recovery purposes) and/or drilling of a much larger number of wells, limiting their storage volumes and increasing the risk of leakage pathways (Van der Velde et al., 2008). Compared to aquifers, because of their long production history, the uncertainty in the geological settings (permeability, heterogeneity, faults, fractures, etc.) of the hydrocarbon fields is relatively low. Moreover, these fields have proven seal and containment integrity and have part of the infrastructure required to handle the gas. The sequestration of $CO_2$ can also be combined with enhancing gas recovery, which could partially compensate for the cost associated with $CO_2$ sequestration (Battashi et al., 2022; Hamza et al., 2021).

However, there are challenges with $CO_2$ storage in depleted gas reservoirs, some of which are related to the thermodynamic properties of $CO_2$. From point of capture (or production) to inside the reservoir pores, $CO_2$ experiences different pressure and temperature conditions and may exist in gas, liquid or, under certain conditions, solid forms. Of particular concern for $CO_2$ storage in low-pressure reservoirs is the so-called Joule-Thomson (J-T) cooling effect, which is caused by expansion of $CO_2$ from high surface pressures to low reservoir pressures.

This is schematically shown by path $A{\rightarrow}D$ in Fig. 1. Point A represents the pressure and the temperature of the transported $CO_2$ at the storage site, which is assumed to be at 100 bar and 30ºC (i.e., $CO_2$ is injected in dense liquid phase). With the assumptions that heat conduction in the wellbore is negligible and that the viscous and gravity forces balance each other, the bottom-hole pressure of $CO_2$ is assumed to be close to that of point *A* (although in practice, the downhole temperature of $CO_2$ will be slightly higher due to heat conduction from the well, and $CO_2$ will most likely be in two phase regime). Isenthalpic expansion of $CO_2$ from 100 bar to a reservoir pressure of 20 bar (point *D*) reduces the $CO_2$ temperature to -10ºC. In general, for reservoirs with pressures lower than 35 bar J-T cooling effect reduces the $CO_2$ temperature to sub-zero values, as illustrated in Fig. 1. This can potentially lead to the formation of hydrates and freezing of pore water, as implied from the pressure-temperature diagram of the water/$CO_2$ system in Fig. 2a. The near-wellbore appearance of hydrates or ice can in turn result in the impairment of well injectivity and/or loss of well-related containment (Vilarrasa & Rutqvist, 2017). Moreover, the thermal stress generated by the cooling effect can lead to fracturing of the rock, the extent of which needs further investigation. An alternative solution will be to heat $CO_2$ before injection (path *A→B→C* in Fig. 1); however, this method is energy (and $CO_2$) intensive and adds further to the cost of $CO_2$ storage. Injection of $CO_2$ in gas form is also not desirable because storing similar mass rates of $CO_2$ require the drilling of multiple wells and requires a larger pore space.

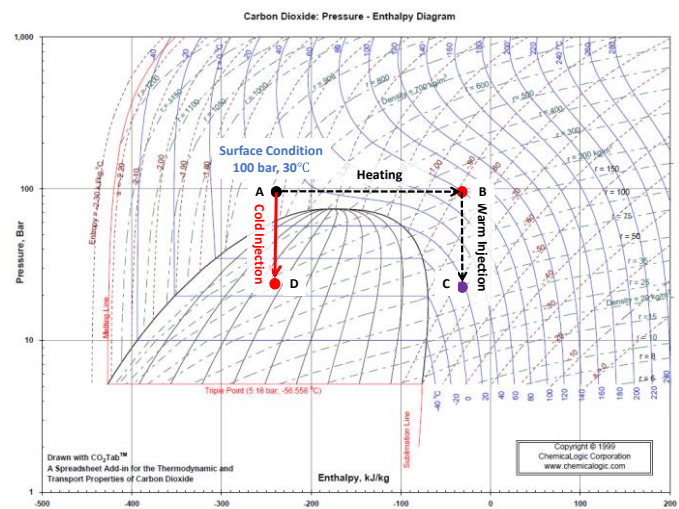

Fig. 1. Pressure-Enthalpy diagram of $CO_2$. The blue colours are the temperature contours.

The long-term and safe storage of $CO_2$ requires a detailed understanding of the physical, chemical, geo-mechanical and thermal effects caused by the injection of $CO_2$ in the reservoirs. This includes mobilisation of fines by CO2-water menisci (Chequer et al., 2020) , salinity alteration (Othman et al., 2019; Parvan et al., 2021; Yu et al., 2018) hydrate formation (Machado et al., 2023) and salt precipitation (Miri et al., 2015; Moghadasi et al., 2004). In addition, to identify the type of risk associated with cold $CO_2$ injection and design a mitigation plan, it is crucial to quantify the range of expected temperatures in the depleted gas reservoirs. Oldenburg (2007) found that for a constant injection rate, lower permeability and higher porosity increase the effect of J-T cooling.

Mathias et al. (2010) derived an exact solution for $CO_2$ injection accounting for J-T cooling into an adiabatic reservoir (ignoring heat exchange with surrounding formations). The analytical model presented in his paper provides explicit formulae for the temperature and pressure profiles, and position of the cold front and quantifies the impact of several parameters on the cooling effect. This model, like all analytical models, allows for fast calculations for multivariant sensitivity studies and can benchmark the numerical methods for more complex modelling (Kacimov & Obnosov, 2023a, 2023b; Katzourakis & Chrysikopoulos, 2019; Moreno et al., 2021). These advantages explain numerous studies on the analytical modelling of problems related to $CO_2$ storage (Ahmadi & Chen, 2019; Celia et al., 2011; Moreno et al., 2021; Ziv Moreno & Avinoam Rabinovich, 2021; Z. Moreno & A. Rabinovich, 2021; Nguyen et al., 2022; Norouzi et al., 2022).

Heat exchange between the reservoir and the adjacent layers can significantly affect the non-isothermal flow in porous media (Bedrikovetsky, 1993; Lawal, 2020). Nevertheless, an exact solution for $CO_2$ injection into low-pressure reservoirs or aquifer accounting for this effect is not available. This paper drives an exact solution accounting for heat exchange between the reservoir and adjacent layers.

The explicit formula allows studying the timely evolution of the temperature profile, including the temperature penetration depth. The solution shows that the temperature profile stabilises with time and the penetration front stops. The solution also allows for calculating the well pressure at the injection well or well injectivity index and the dynamics of the heat penetration depth The explicit formulae for temperature and pressure allow calculating maximum injection rate that avoid formation of hydrates, i.e. the path well-reservoir in phase diagram does not enter the hydrate domain (Fig. 2). The validity of the model with Newton's law for heat exchange is determined by comparison with the exact solution.

The structure of the paper is as follows. Section 2 describes the major model assumptions and presents the mathematical model and the ensuing equations. Section 3 presents the explicit solution of the model, derived in Appendix A. Section 4 provides the calculation for the area of validity of the steady-state heat exchange model. Appendix B presents the exact solution for heat flux in adjacent layers and derives the criterion used to establish the domain of validity for the steady-state heat exchange term used in this paper. Section 5 presents the results of the analytical model. Section 6 discusses the sensitivity study of the model, introducing the impact of dimensionless parameters on the temperature.

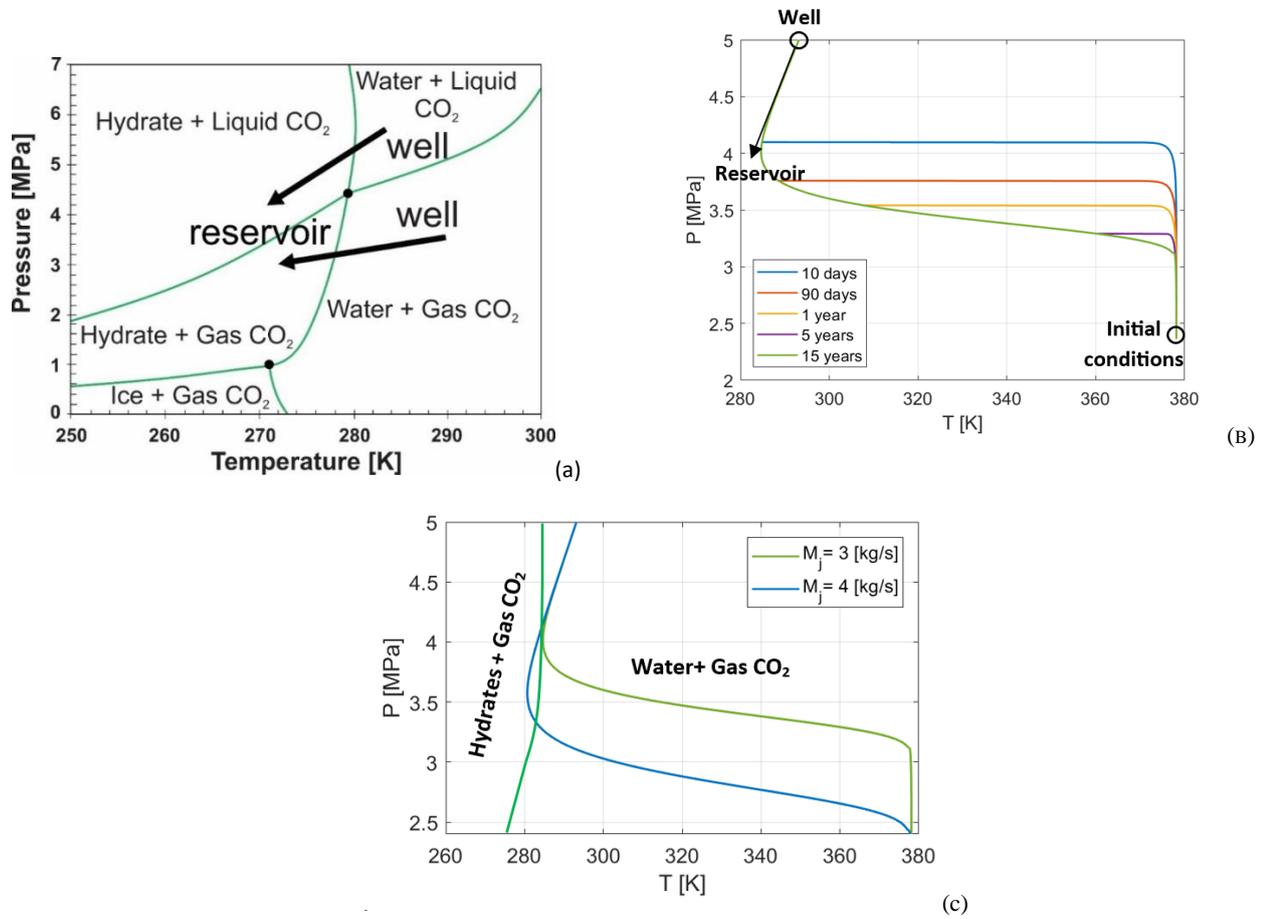

Fig. 2. (a) Pressure-Temperature phase diagram of water-$CO_2$ binary system, reproduced from Voronov et al. (2016), (b) (P,T) diagram based on our model. (c) Mass injection rate effect on a (P,T) diagram, at time $t_D = 5.4$.

## 2. MATHEMATICAL MODEL

This section presents the mathematical model for temperature profile evolution during $CO_2$ injection: the model assumptions (section 2.1) and derivation of the heat-transfer equation (section 2.2).

2.1 Model Assumptions

The main assumptions of the model are: (i) 1D radial unsteady-state single-phase non-isothermal flow in an infinite homogeneous reservoir; (ii) water, $CO_2$, and rock are incompressible; (iii) the reservoir (with permeability $k$ [L$^2$] and porosity $\phi$) contains water with an initial water saturation $S=1$; piston-like displacement of water by $CO_2$ resides constant saturation $S_{wi}$ of immobile water behind the displacement front; (iv) temperature at outer boundaries of under and overburden shales is equal to initial temperature $T_I$ [K]; (v) the temperature of the injected $CO_2$ in the wellbore is constant $T_J$ [K]; (vi) Joule-Thomson coefficient, permeability, porosity, heat conductivity and heat capacity of rock and fluid are constant; (vii) $CO_2$ is injected at a constant rate of $q_J$ [M$^3$T$^{-1}$], and pressure at drainage radius is constant; (viii) water evaporation into gaseous phase is neglected; (ix) density and viscosity of fluids are constant; (x) heat exchange between surrounding shales and the reservoir occurs under steady-state conditions.

The model assumes steady-state heat exchange between the reservoir and the surrounding formations (Newton's law) $2\pi r \gamma^s l^{-1}(T - T_I)$ (Jang et al., 2022). So, the reservoir is bounded by overburden and underburden shales with thickness $l$ [L]. Here, $r$ [L] is the radial distance from the injection well, $T(r,t)$ [K] is the fluid temperature in the reservoir, $\gamma^s$ [MLT$^{-3}$K$^{-1}$] is the heat-transfer coefficient (or overall thermal conductivity) of the seal in contact with $CO_2$, and $T_I$ [K] is the initial temperature of the reservoir. This assumption is valid when the reservoir is hydrodynamically separated above and below from the adjacent layers by the impermeable seals. The heat fluxes behind and ahead of the horizontal boundary between the adjacent formations and seals are assumed equal. Heat flux is a product of thermal conductivity and temperature gradient. The thermal conductivity of an impermeable seal, e.g., shale is determined by that of the solid matrix, which is significantly lower than the thermal conductivity of the water residing in the adjacent permeable formation and forming a conductive cluster. Therefore, the temperature gradient in shales is expected to be significantly lower than the temperature gradient in the adjacent formation. This supports the assumption that the temperature at the upper and lower bounds of the adjacent shales layers remains equal to the initial temperature $T_I$.

Newton's law for the heat exchange between the reservoir and the adjacent layers is a commonly-used assumption for 1D non-isothermal flows in porous media (Atkinson & Ramey, 1977; Batycky & Brenner, 1997; Fedorov & Sharafutdinov, 1989; Gordeev et al., 1987; Jang et al., 2022; LaForce et al., 2014; Muradov & Davies, 2012; Muradov & Davies, 2009; Payne & Straughan, 1998; Pires et al., 2006; Xu et al., 2013; Yortsos & Gavalas, 1982; Zazovskii, 1983; Zolotukhin, 1979). However, the physical-geological conditions for Newton's law validity are not present in the literature. Derivations in Appendix B present the dimensionless criteria determining the validity of Newton's law. Furthermore, it is assumed that the heat exchange between the injected $CO_2$, water, and the rock occurs instantly. The heat transfer due to conduction in the flow direction is ignored compared to advection. The schematic of the radial problem for $CO_2$ injection into a depleted reservoir is given in Fig. 3.

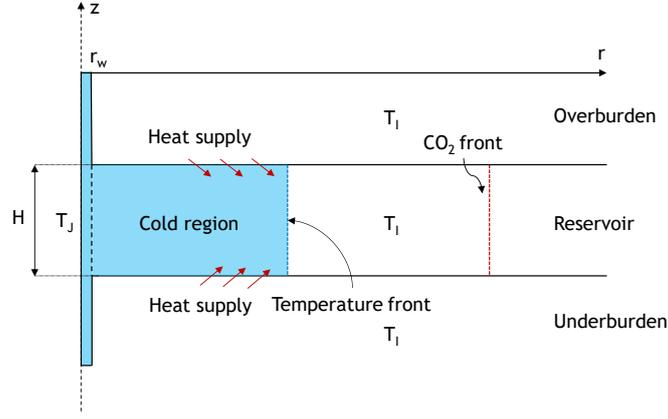

Fig. 3. Schematic of $CO_2$ injection into a depleted reservoir. The cold part (due to Joule-Thomson cooling effect) is shown in blue.

2.2 1-D Flow Equations

The energy-balance equation consists of the accumulation term of heat in the rock, $CO_2$, and water, advective heat transport, the J-T effect, and heat exchange of the reservoir with the adjacent layers (Lake et al. 2014; Wang 2022; Hashish & Zeidouni 2022):

$$2\pi r H \frac{\partial}{\partial t}\{[\phi(1-S_{wi})\rho_f c_f + \phi S_{wi}\rho_w c_w + (1-\phi)\rho_s c_s]T\} + q_J \rho_f c_f \left[\frac{\partial T}{\partial r} - \alpha_{JT}\frac{\partial p}{\partial r}\right] = -2\pi r \frac{\gamma^s}{l}(T-T_I) \quad (1)$$

where $\rho_f$ [ML$^{-3}$] is density of $CO_2$, $c_f$ [L$^2$T$^{-2}$K$^{-1}$] is the specific heat capacity of $CO_2$, $\rho_w$ [ML$^{-3}$] is the density of water, $c_w$ [L$^2$T$^{-2}$K$^{-1}$] is the specific heat capacity of water, $\rho_s$ [ML$^{-3}$] is the rock density, $c_s$ [L$^2$T$^{-2}$K$^{-1}$] is the rock specific heat capacity, $q_J$ [L$^3$T$^{-1}$] is the injection flow rate, $H$ [L] is the formation thickness, $\alpha_{JT}$ [M$^{-1}$LT$^2$K] is the J-T coefficient, and $p$ [ML$^{-1}$T$^{-2}$] is the fluid pressure. The unknown in Eq. (1) is the reservoir temperature $T(r,t)$.

The J-T coefficient of $CO_2$, $\alpha_{JT}$, depends on the pressure and temperature as shown in Fig. 4. For the depleted reservoirs with pressures of less than 5 MPa, $\alpha_{JT}$ increases with a decreasing temperature. For a fixed temperature, $\alpha_{JT}$ can be assumed independent of pressure for these reservoirs, even though for larger pressures $\alpha_{JT}$ decreases with increasing pressure.

The overall heat transfer coefficient $\kappa$ [MT$^{-3}$K$^{-1}$] is defined as (Lawal, 2020):

$$\kappa = \gamma^s l^{-1} \quad (2)$$

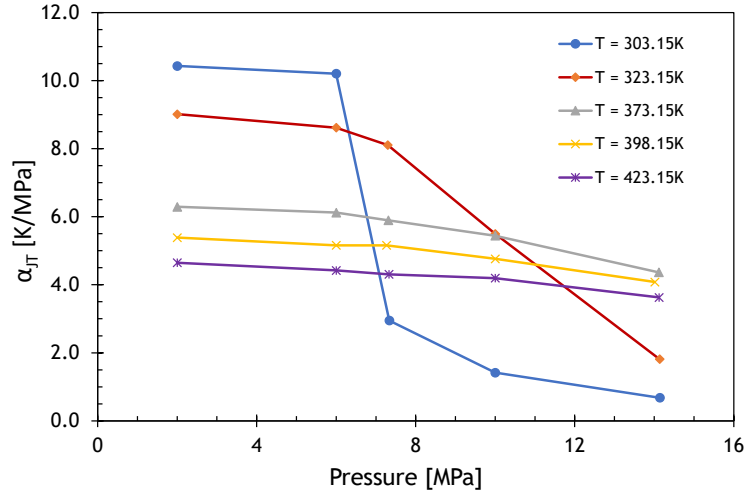

Fig. 4. Joule-Thomson coefficient of $CO_2$ as a function of pressure for different temperatures (modified from Gao et al. (2021) )

It has been experimentally observed that when a fluid with a different temperature than the temperature of the porous medium is injected the heat loss (or gain) per unit length is almost constant. Consequently, the overall heat transfer coefficient, $\kappa$, is assumed constant in this study. In other words, the heat conductivity in the *z*-direction is assumed to be very large, which results in a constant temperature across the vertical thickness of the reservoir (LaForce et al., 2014). However, $\kappa$ may vary with injection rate, the surface area in contact with surroundings, time, thermal conductivity of the surrounding porous medium, and/or the heat capacity of the reservoir, and the type of process (cold or hot fluid injection) (LaForce et al., 2014). Cold $CO_2$ injection into a reservoir filled with a hot fluid is analogous to cold water injection in geothermal reservoirs. The experimental data show that for cold water injection lower injection rates are more advantageous for heat extraction from the surrounding rocks.

The volumetric injection rate, $q_J$, is defined as:

$$q_J = M_J \rho_f^{-1} \qquad (3)$$

where $M_J$ [MT$^{-1}$] is the mass injection rate. The position of the $CO_2$ front $r_f$ [L] is calculated assuming gas incompressibility:

$$M_J t = \pi(r_f^2(t) - r_w^2)(1 - S_{wi})\phi H \rho_f \qquad (4)$$

$$r_f = \left(r_w^2 + \frac{M_J}{\pi \phi H \rho_f (1 - S_{wi})} t\right)^{1/2} \qquad (5)$$

The pressure gradient in the reservoir is calculated using Darcy's law:

$$\frac{\partial p}{\partial r} = -\frac{\mu_f q_J}{2\pi r H k_{rf}^e k} \tag{6}$$

where $\mu_f$ [ML$^{-1}$T$^{-1}$] is viscosity of CO$_2$, and $k_{rf}^e$ is the end-point relative permeability of CO$_2$. Finally, the substitution of Eq. (6) into Eq. (1) yields the first order partial differential equation:

$$\frac{\partial}{\partial t}\{[\phi(1-S_{wi})\rho_f c_f + \phi S_{wi}\rho_w c_w + (1-\phi)\rho_s c_s]T\} + \frac{q_J \rho_f c_f}{2\pi r H}\frac{\partial T}{\partial r} + \alpha_{JT}\frac{\mu_f \rho_f c_f q_J^2}{4\pi^2 r^2 H^2 k_{rf}^e k} = -\frac{\kappa}{H}(T-T_I) \tag{7}$$

The boundary condition (BC) correspond to the injection of CO$_2$ with a constant pressure and temperature:

$$r = r_w: \quad T = T_J, \quad p = p_J \tag{8}$$

where $r_w$ [L] is the well radius.

The initial condition (IC) corresponds to the equality of the initial temperatures of the main reservoir rock and the adjacent layers, i.e.,

$$t = 0: \quad T = T_I \tag{9}$$

The following dimensionless variables are introduced to nondimensionalize Eq. (7).

$$t_D = \frac{\gamma^s t}{H l c^{res}}, \quad r_D = \frac{r}{r_w}, \quad T_D = \frac{T}{T_I} \tag{10}$$

Further, the dimensionless constants $V^*$ and $A^*$ are defined as:

$$V^* = \frac{q_J \rho_f c_f l}{2\pi r_w^2 \gamma^s} \tag{11}$$

$$A^* = V^* \frac{\alpha_{JT}\mu_f q_J}{2\pi k_{rf}^e k H T_I} \tag{12}$$

By substituting Eqs. (10-12), the dimensional equation Eq. (7) transforms into the dimensionless form:

$$\frac{\partial T_D}{\partial t_D} + V^*\frac{1}{r_D}\frac{\partial T_D}{\partial r_D} = -A^*\frac{1}{r_D^2} - (T_{r,D} - 1) \tag{13}$$

Additionally, the BC Eq. (8) and IC Eq. (9) are also expressed in dimensionless form as:

$$t_D = 0: T_D = 1 \tag{14}$$

$$r_D = 1: T_D = \frac{T_J}{T_I} \tag{15}$$

It is assumed that the heat transfer within the rock occurs instantaneously, i.e., the injected cold $CO_2$ is instantly heated up by the residing water and rock. This implies that the temperature front lags significantly behind the $CO_2$ front inside the reservoir.

## 3. EXACT ANALYTICAL SOLUTION

The exact solution of Eq. (13) with initial and boundary conditions Eqs. (14-15) is obtained by the method of characteristics (Boyce & DiPrima, 2013; Polyanin & Zaitsev, 2003) derived in Appendix A. The substitution of the dimensionless parameters given by Eqs. (10-12) into Eq. (A16) yields the dimensional form of the solution as:

$$T(r,t) = \begin{cases} T_I + (T_J - T_I)e^{-\frac{B}{V}\pi(r^2 - r_w^2)} + \frac{\alpha_{JT}\mu_f q_J}{4\pi k_{rf}^e kH}\left[Ei\left(\frac{B}{V}\pi r_w^2\right) - Ei\left(\frac{B}{V}\pi r^2\right)\right]e^{-\frac{B}{V}\pi r^2}; & \pi r^2 < \pi r_w^2 + Vq_J t \\ T_I + \frac{\alpha_{JT}\mu_f q_J}{4\pi k_{rf}^e kH}\left\{Ei\left(\frac{B}{V}(\pi r^2 - Vq_J t)\right) - Ei\left(\frac{B}{V}\pi r^2\right)\right\}e^{-\frac{B}{V}\pi r^2}; & \pi r^2 > \pi r_w^2 + Vq_J t \end{cases} \quad (16)$$

Here, the dimensional constants $V$ and $B$ are:

$$V = \frac{\rho_f c_f}{H[\phi(1-S_{wi})\rho_f c_f + \phi S_{wi}\rho_w c_w + (1-\phi)\rho_s c_s]} \quad (17)$$

$$B = \frac{\kappa}{q_J H[\phi(1-S_{wi})\rho_f c_f + \phi S_{wi}\rho_w c_w + (1-\phi)\rho_s c_s]} \quad (18)$$

The analytical model (A16), (16) can be extended for the case where the injected temperature changes with time – $T_J(t_D)$. In this case, $T_J$ in Eq. (A16) must be substituted by $T_J(t_D-(r_D^2-1)/2V^*)$.

Let us define the depth of temperature wave penetration as a coordinate of the "centre" of the wave. For variable $r$, the probability distribution function can be defined as:

$$F(r) = \frac{(T(r,t) - T_I)r}{\int_{r_w}^{\infty}(T(r,t) - T_I)r\, dr} \quad (19)$$

The arithmetic mean of $r$ is

$$r_p = \int_{r_w}^{\infty} F(r)r\, dr = \frac{\int_{r_w}^{\infty}(T(r,t) - T_I)r^2\, dr}{\int_{r_w}^{\infty}(T(r,t) - T_I)r\, dr} \quad (20)$$

## 4. AREA OF VALIDITY OF THE MODEL

This section determines the area of validity of the analytical model, including comparison between the solutions for adiabatic and non-adiabatic reservoirs (section 4.1), formulation of the method of 1D model validation by comparison with 2D solution (section 4.2), and calculations of validity intervals for the model parameters and independent variables (section 4.3).

4.1 Definition of the Basic Case for Sensitivity and Model Validation

To validate the developed analytical model, the solution given by Eq. (19) was compared with the results of Mathias et al. (2010) in the limit of no heat transfer between the reservoir and the adjacent layers, i.e., $\kappa \to 0$. The parameters provided in Table 1 are used as a basic case for all calculations in this paper. Table 2 presents the intervals of variation of variables. Fig. 5 shows the close agreement between the numerical implementations of two models. The time intervals in Fig. 5 correspond to dimensionless time, $t_D$ = 0.01, 0.02, 0.09, 0.3, 1.8, 5.4, 7.1.

As expected, the temperature front falls behind the $CO_2$ front because of the heat provided by the rock (the volumetric heat capacity of the rock is much larger than those of both water and $CO_2$). The temperature decreases from injection temperature to a minimum value and then sharply increases to the initial temperature of the reservoir. There are two distinctive features for the solution of the case, where $\kappa \to 0$: (1) temperature decreases to a minimum value just upstream of the sharp front at each timestep, and (2) the temperature front keeps moving inside the reservoir for as long as $CO_2$ injection continues. The sharp increase to the initial temperature (shock front) is due to absence of conduction in the analytical model. In accordance with the properties of the radial flow (i.e., large pressure drops near the injection point), the cold front penetrates quickly inside the reservoir. If the injected $CO_2$ is undersaturated, the residing water in the pores evaporates; and therefore, the water saturation reduces to near-zero close to the injection well. However, the speed of the temperature front is expected to be larger than that of the drying front such that a region with a non-zero water saturation appears between the dry-out and cold fronts in a relatively short time after injection of $CO_2$. This is schematically shown in Fig. 6. If the pressure and temperature fall within in the hydrate region on the $CO_2$/water phase diagram in Fig. 2, $CO_2$ hydrates might form potentially leading to blockage of $CO_2$ flow.

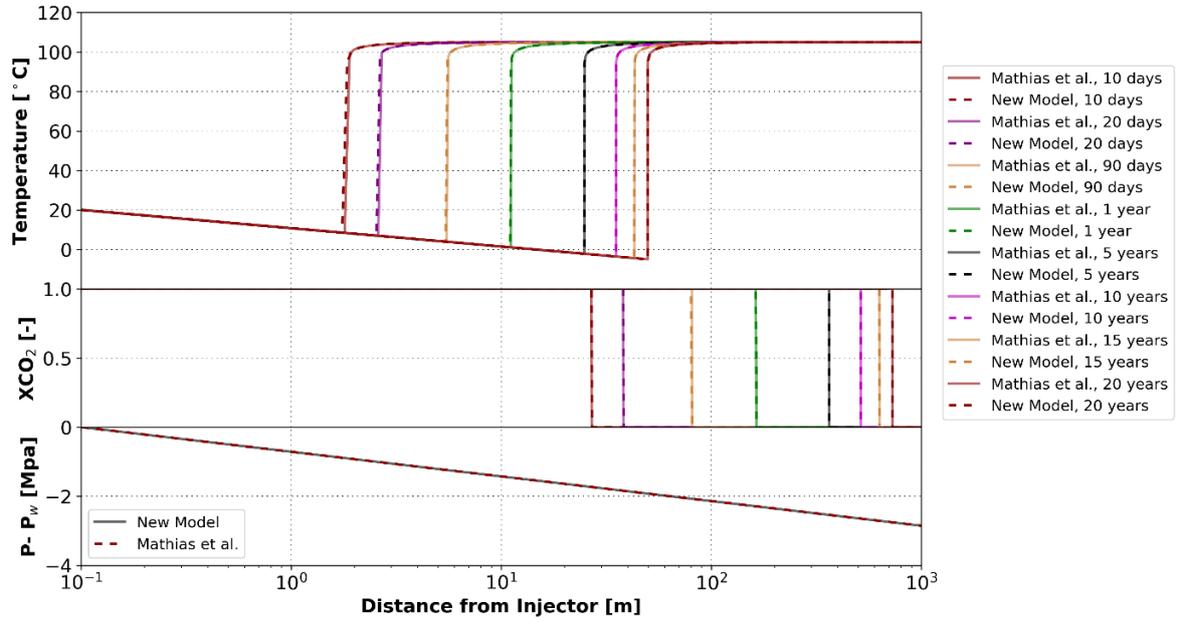

Fig. 5. Comparison of the results of the current model with Mathias et al. (2010) model ($\kappa = 0$).

Table 1. Model parameters used in this study.

| Parameter | Symbol | Unit | Value |
|---|---|---|---|
| Formation thickness | $H$ | m | 91 |
| Porosity | $\phi$ | [-] | 0.11 |
| Permeability | $k$ | m$^2$ | $2 \times 10^{-15}$ |
| End point relative permeability | $k_{rf}^e$ | [-] | 1 |
| Well radius | $r_w$ | m | 0.1 |
| Rock density | $\rho_S$ | kg/m$^3$ | 2600 |
| Water density | $\rho_w$ | kg/m$^3$ | 992 |
| CO$_2$ density | $\rho_f$ | kg/m$^3$ | 141.4 |
| CO$_2$ viscosity | $\mu_f$ | Pa.s | $16.7 \times 10^{-5}$ |
| Heat capacity of rock | $c_S$ | J/kg/K | 1000 |
| Heat capacity of water | $c_w$ | J/kg/K | 4037 |
| Heat capacity of CO$_2$ | $c_f$ | J/kg/K | 904 |
| Joule Thomson coefficient | $\alpha_{JT}$ | K/Pa | $10.2 \times 10^{-6}$ |
| Reservoir pressure | $p_w$ | Pa | $50 \times 10^5$ |
| Reservoir temperature | $T_I$ | K | 378.15 |

| Injection temperature | $T_J$ | K | 293.15 |
|---|---|---|---|
| $CO_2$ mass injection rate | $M_J$ | kg/s | 3 |
| Connate water saturation | $S_{wi}$ | [-] | 0.2 |

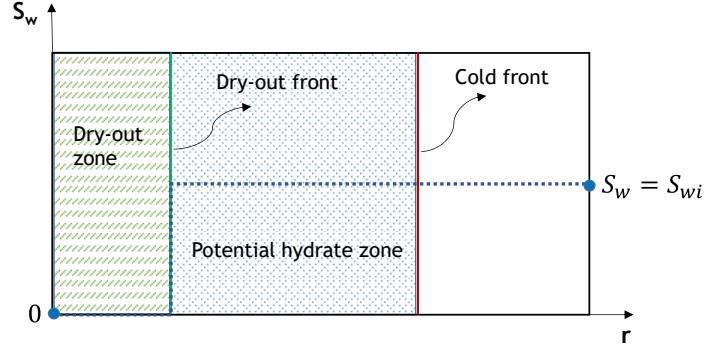

Fig. 6. Formation of a region with a non-zero water saturation between dry-out and cold fronts.

4.2 Formulating the Criterion for Validity of the Heat Exchange Term.

The primary assumption of the model given by Eq. (1) is steady state heat exchange between surrounding formations and impermeable shale that confines the reservoir, which is the Newton's law. This section establishes the domain of validity for the heat exchange term used in the energy balance (Eq. (1)), aiming to determine the condition under which the discrepancy between the temperature at the interface, $\theta(z_D = 0, t_D)$, and the initial reservoir temperature, $T = T_I \rightarrow \theta = 1$, is considered insignificant. In our case, we constrain the discrepancy between the two solutions for temperature to the difference of less than or equal to 10%.

We utilize the solutions obtained in Appendix B for the temperature profile at the interface between adjacent impermeable shale, as derived in Eq. (B19). The condition in dimensionless variables is as follows:

$$1 - \theta(z_D = 0, t_D) < \varepsilon, \quad \varepsilon = 0.1 \tag{21}$$

In this analysis we introduce two dimensionless constants, $A^*$ and $V^*$ given in Eq. (11) and Eq. (12). Also, we present the constant $\omega$ determined in Appendix B as Eq. (B9):

$$\omega = \sqrt{\frac{c^{res}\gamma^s H}{c^a \gamma^a l}} \tag{22}$$

Newton's law is valid when the temperature on the outer bounds of the adjacent seals is equal to the initial reservoir temperature. This condition depends on conductivity and capacity of the adjacent formation layers i.e., depends on the heat conductivity ($\gamma^a$) and volumetric heat capacity ($c^a$). Indeed, equality of the temperature on the outer shale bounds to initial reservoir temperature corresponds to $\omega$ tending to zero.

The intervals for parameters $A^*$, $V^*$ and $\omega$ are determined from typical fluid and rock properties. The details below highlight the values adopted in this work. The heat capacities of gas, water and rock varying from 709 J (kg K)$^{-1}$ to 1476 J (kg K)$^{-1}$, 3965 J (kg K)$^{-1}$ to 4335 J (kg K)$^{-1}$ and from 776 J (kg K)$^{-1}$ to 1215 J (kg K)$^{-1}$, respectively. The heat conductivities of shale and adjacent formations ranges from 1 W (m K)$^{-1}$ to 3.7 W (m K)$^{-1}$ and from 3.7 W (m K)$^{-1}$ to 5 W (m K)$^{-1}$, respectively. The densities of gas, water and rock vary from 0.5 kg m$^{-3}$ to 1236 kg m$^{-3}$, from 527 kg m$^{-3}$ to 1000 kg m$^{-3}$ and from 2270 kg m$^{-3}$ to 3200 kg m$^{-3}$, respectively. In addition, the thickness of the reservoir and the shale vary from 1 m to 100 m and from 1 m to 3.33 m, respectively. Well radius varies from 0.05 m to 0.13 m. Injection rate varies from 0.005 m$^3$ s$^{-1}$ to 0.04 m$^3$ s$^{-1}$. J-T coefficient varies from $7\times10^{-6}$ K/Pa to $1.1\times10^{-5}$ K/Pa. Additionally, CO$_2$ viscosity varies from $1.1\times10^{-5}$ Pa s to $2.7\times10^{-4}$ Pa s. Permeability varies from $3.9\times10^{-16}$ m$^2$ to $5.9\times10^{-14}$ m$^2$. Finally, porosity varies from 0.1 to 0.3 and connate water saturation varies from 0.2 to 0.4.

Given that our solution in Eq. (B19) is notably influenced by system parameters- $A^*$, $V^*$ and $\omega$, a domain made of the three dimensionless constants, ($A^*$, $V^*$, $\omega$) in Eq. (B19) is established. The eight vertices of the tetrahedron represented by dots in Fig. 7 correspond to the extreme values of $A^*$, $V^*$ and $\omega$. The inner dot numbered 9 corresponds to the unperturbed base case given in Table 1**Error! Reference source not found.**.

Fig. 7. A tetrahedron defining the domain of the three dimensionless constants in the system.

Table 2 summarizes the critical values of the 9 coordinate points illustrated in Fig. 7.

Table 2. Critical points of the ($A^*$, $V^*$, $\omega$) domain

| Vertices | $V^*$ | $A^*$ | $\omega$ |
|---|---|---|---|
| 1 | 4.51 | $1.08\times10^{-4}$ | $1.56\times10^{-1}$ |
| 2 | 4.51 | $2.14\times10^{3}$ | $1.56\times10^{-1}$ |
| 3 | 4.51 | 2.14×10³ | 1.48×10¹ |
| 4 | 4.51 | 1.08×10⁻⁴ | 1.48×10¹ |
| 5 | 1.55×10⁷ | 1.08×10⁻⁴ | 1.56×10⁻¹ |
| 6 | 1.55×10⁷ | 2.14×10³ | 1.56×10⁻¹ |
| 7 | 1.55×10⁷ | 2.14×10³ | 1.48×10¹ |
| 8 | 1.55×10⁷ | 1.08×10⁻⁴ | 1.48×10¹ |
| 9 | 1.63×10⁴ | 1.28×10² | 9.29 |

4.3 Determining the Area of Validity

Our aim is to determine a region of validity where the condition in Eq. (21) is upheld. To visualize this, we plot $1 - \theta(z_D = 0, t_D)$ against dimensionless time for various radial distances in the reservoir in Fig 8a. We also obtain a generalized area of validity in the ($r_D$, $t_D$) plane in Fig 8b, where $r_c$ is the critical radius.

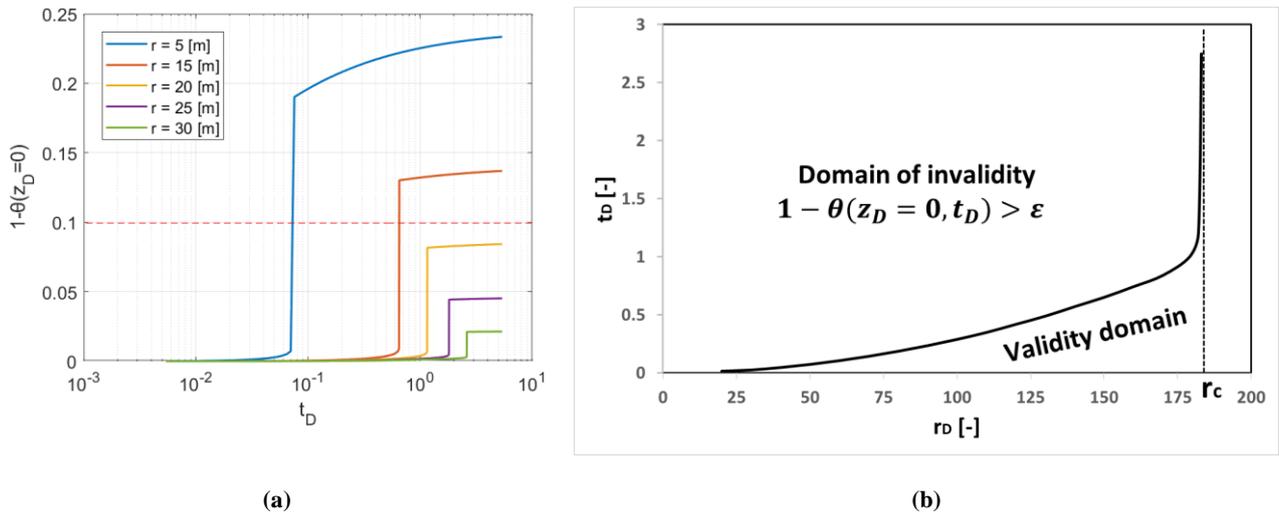

(a)          (b)

Fig 8. Validity of steady state heat exchange (a) validity vs. time for different radii. (b) domain of validity

To explore the scope of the area of validity, we show the impact of five typical values of $\omega$, at a constant radial distance of $r = 5$ m, in Fig. 9.

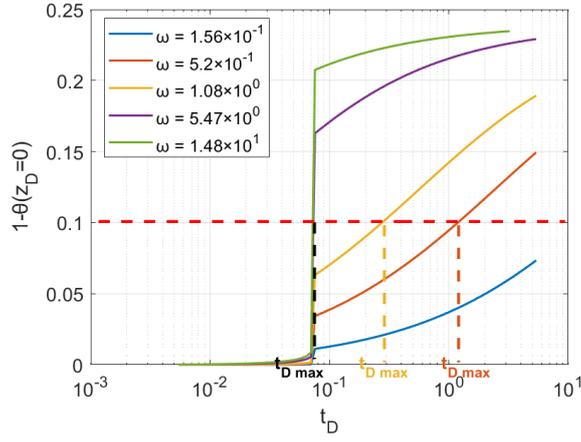

Fig. 9. Newton's law criterion vs. time, for different $\omega$ values. Plotted for $r = 5$ m.

The 10% error dashed line in Fig 8a and Fig. 9 indicates our chosen criterion for Newton's law, i.e., $1 - \theta(z_D = 0, t_D) < 0.1$. In case a different criterion is needed, the area of validity for Newton's law changes accordingly. To provide an analysis on the validity in each of the points presented in Table 2, we introduce the definition for radius of injection, following Eq. (5):

$$r_{inj} = \sqrt{\frac{M_J}{\pi \phi H \rho_f (1 - S_{wi})} t_{inj}} \tag{23}$$

Here, $t_{inj}$ [T] is the time of injection. Considering $t_{inj} = 10$ years, we plot $1 - \theta(z_D = 0, t_D)$ for each point in Table 2 under the constraint of the chosen 10% error margin. The plots in Fig 8a show that for short distances near the well ($r = 5$ m , $r = 15$m) validity of the model based on Newton's law is limited. Newton's law is obeyed until the moment of arrival of temperature front to these near wellbore points, beyond which invalidity ensues. However, for larger distances away from the wellbore, the solution is valid for the overall time domain. The domain of validity is further shown in Fig 8b. There exists such a critical radius ($r_c$) beyond which the solution is valid for all times.

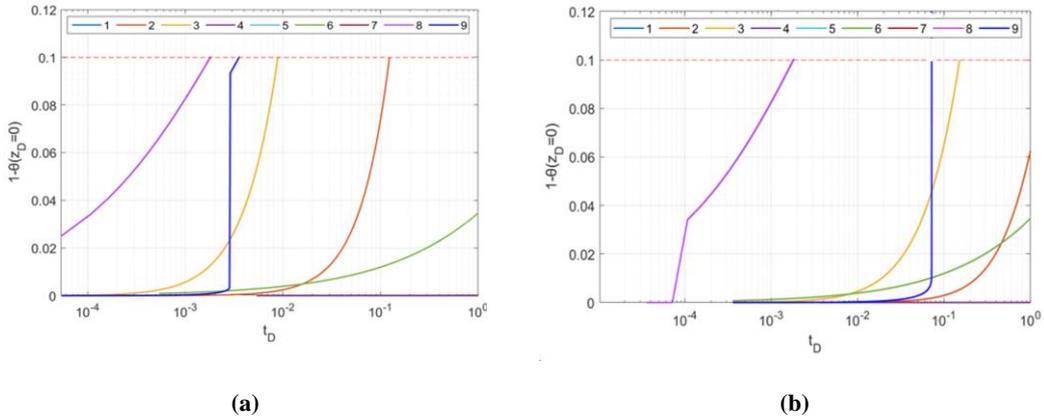

(a)    (b)

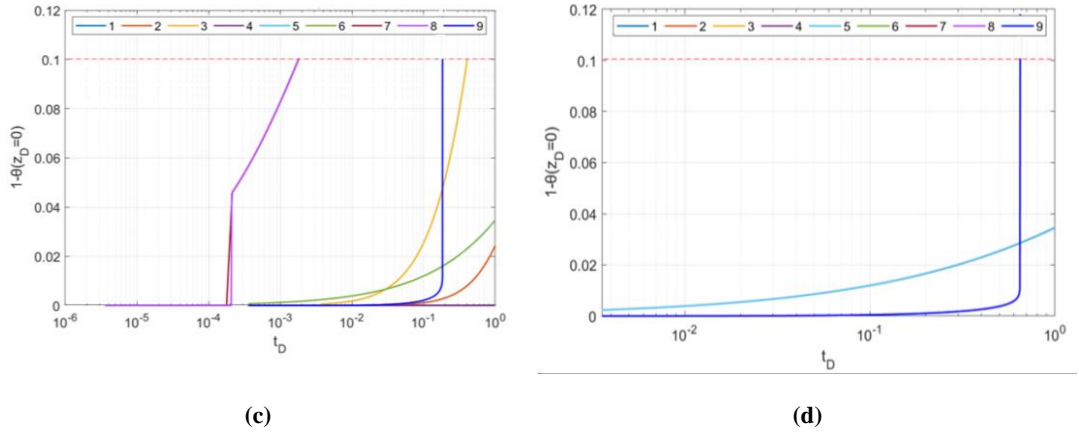

| (c) | (d) |

Fig. 10. Difference between temperature on the outer bounds of shale and initial reservoir temperature, to validate Newton's law, temperature history at the distance from injectors at (a) $r = 0.002\ r_{inj}$, (b) $r = 0.01\ r_{inj}$, (c) $r = 0.015\ r_{inj}$, (d) $r = 0.03\ r_{inj}$

## 5. RESULTS OF THE ANALYTICAL MODELLING

This section presents the results of the derived analytical model and discusses their relevance for $CO_2$ storage in the low-pressure or depleted fields. The impact of heat exchange between the main formation and the surrounding rock on the propagation of the temperature front is shown in Fig. 11. A heat transfer coefficient of $\kappa = 2.5$ W/m²/K was used in the calculations (Cermak & Rybach, 1982; Labus & Labus, 2018; Robertson, 1988; Schoen, 2015). For the purpose of comparison, Fig. 11 further shows the case of no heat exchange ($\kappa \to 0$) with the dashed curves. Initially and for relatively short times, the impact of the heat gain from the surrounding layers is not significant.

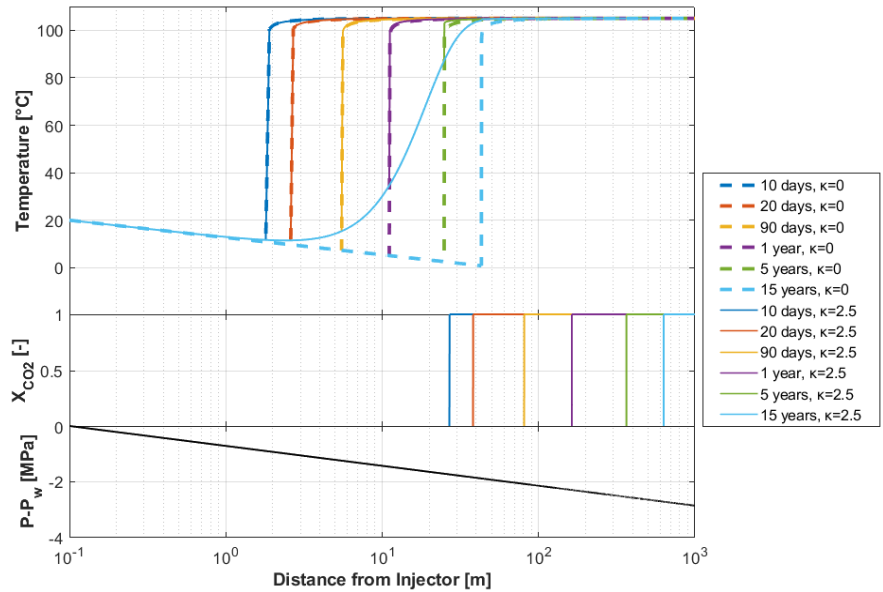

Fig. 11. The impact of heat exchange on temperature profile (solid lines) in comparison to adiabatic system (dashed lines), $CO_2$ front position and Pressure drop for varying times.

However, as $CO_2$ injection continues, the heat supplied by these layers heats the injected $CO_2$ and therefore the temperature increases slightly before jumping to the initial temperature of the reservoir. Unlike the case with no heat exchange, the temperature upstream of the shock is not the lowest temperature. Indeed, a key difference is that the position and the value of the minimum temperature remains fixed for all times in the new model. The area between the solid and the dashed curves in Fig. 11 represents the extent of the impact of the heat exchange. The area increases with time and as a result the length of the shock front becomes smaller every timestep, until it disappears completely. This corresponds to the time at which the system reaches steady state, and the temperature front no longer moves. The solution of the ordinary differential equation (ODE) is obtained by the separation of variables:

$$T(r) = T_I + (T_J - T_I)e^{-C(\pi r^2 - \pi r_w^2)} + D[Ei(C\pi r_w^2) - Ei(C\pi r^2)]e^{-C\pi r^2} \qquad (24)$$

Here, $C$ and $D$ are defined as follows:

$$C = \frac{\kappa}{q_J \rho_f c_f}, \quad D = \frac{\alpha_{JT} \mu_f q_J}{4\pi k_{rf}^e k} \qquad (25)$$

The time intervals in Fig. 11 correspond to dimensionless timey, $t_D$ = 0.01, 0.09, 0.3, 1.8, and 5.4, respectively. Eqs. (24) and (25) indicate that the position of the steady state temperature depends largely on the injection rate, the volumetric heat capacity of $CO_2$ and rock properties. The pressure profile under the assumptions of gas incompressibility and constant viscosity is steady state; it is obtained by integration of Eq. (6) with BC Eq. (8) for pressure. The temperature profile in Fig. 11 shows temperature decreases from the well to the temperature front, and then monotonically increases until initial temperature. Minimum temperature is achieved exactly behind the temperature front. It is crucial to determine a minimum injection rate at which the temperature behind the temperature front and pressure along the temperature front are not conducive for hydrate formation (as illustrated by the phase diagram in Fig. 2a). Additionally, Fig. 2b. shows the pressure-temperature path in the phase diagram, where the reservoir point corresponds to minimum temperature. The length of the segment, injection well-reservoir, as it follows from Eq. (6), is proportional to injection rate $q_J$. This allows determining the minimum rate of hydrate formation. Fig. 2c shows two solutions that corresponds to mass injection rates of $M_J = 3$ kg/s and $M_J = 4$ kg/s.

According to Fig. 2c, hydrates formation occurs when the plot crosses the Hydrates + Gas $CO_2$ phase boundary, which is evident only for the higher injection rate.

## 6. SENSITIVITY STUDY OF THE MODEL

This section investigates the sensitivity of the analytical model to the model parameters: heat transfer coefficient (subsection 6.1), permeability (subsection 6.2), and reservoir thickness as well as injection rate (subsection 6.3). Additionally, the effect of the dimensionless parameters $A^*$ and $V^*$ on the temperature profile is investigated in subsection 6.4.

### 6.1 Effect of Heat Transfer Coefficient

The impact of the heat transfer coefficient, $\kappa$, is shown in Fig. 12. With the increase of $\kappa$, a higher amount $CO_2$ is heated and therefore the cold front is retarded, compared to the case where $\kappa$ is smaller. Moreover, the minimum temperature is slightly lower for the case with the smaller $\kappa$, as illustrated in Fig. 12a.

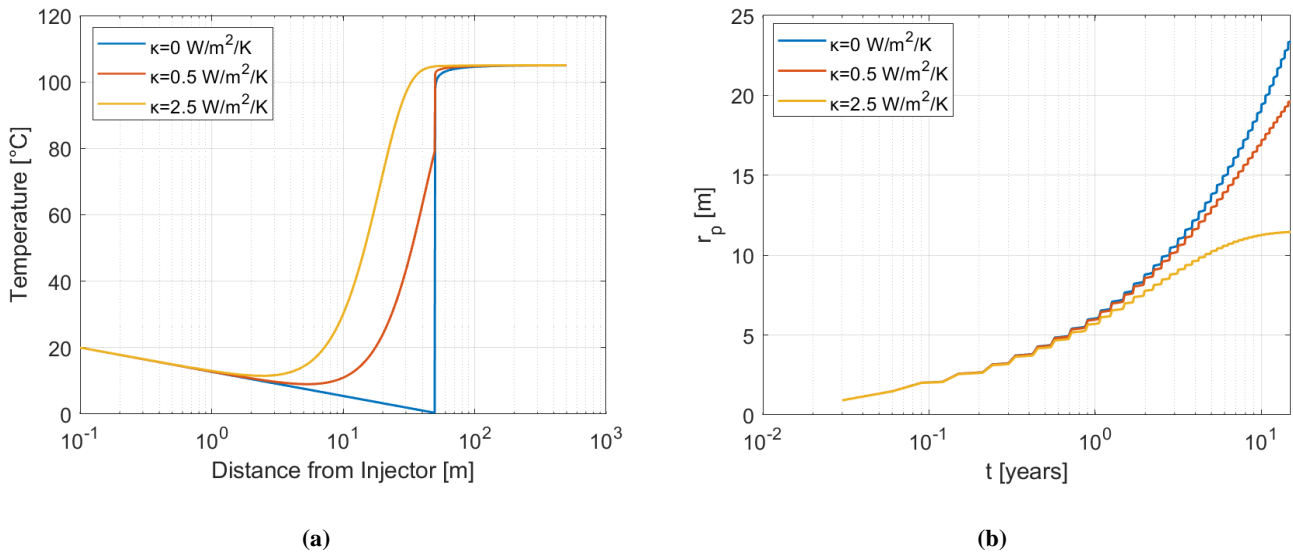

(a)         (b)

Fig. 12. (a) The impact of the heat transfer coefficient, $\kappa$, on the temperature profile. (b). The impact of the heat transfer coefficient, $\kappa$, on the temperature penetration depth.

Fig. 12b shows that in the adiabatic reservoir, the penetration depth, which was introduced by Eq. (20), does not stabilize (blue curve). Stabilization of the heat penetration is achieved with the presence of heat exchange in the system. For the given temporal range, the curve corresponding to the lower value for heat conductivity does not stabilize, while the curve corresponding to the higher value achieves stabilization. The three values of heat conductivity presented in Fig. 12 correspond to the following dimensionless parameters- $(A^*,V^*\to\infty)$, $(A^*=7.2\times10^2, V^*=8.6\times10^4)$, $(A^*=1.4\times10^2, V^*=1.7\times10^4)$, respectively.

6.2 Effect of Permeability

Fig. 13 shows the sensitivity of the model to varying rock permeabilities. As stated earlier, the magnitude of temperature reduction due to the J-T effect depends on the pressure gradient and the J-T coefficient. For fixed injection pressure and flowrate at the boundary, the pressure gradient increases with decreasing reservoir permeability, causing an increased temperature drop. However, this conclusion is valid only when equal flowrates are considered for different permeabilities. For a heterogeneous reservoir, the injected $CO_2$ will be distributed in proportion to the permeability of each layer. If the permeability contrast is large, then the large fraction of $CO_2$ will flow through the high permeability layer causing a larger pressure drop compared to the low-permeability layer. Therefore, for heterogeneous reservoirs with large permeability contrast, the "isenthalpic" expansion of $CO_2$ will result in lower temperatures in the high permeability layers.

It is worth mentioning that when $CO_2$ becomes liquid or the pressure exceeds the critical pressure of $CO_2$ the magnitude of $\alpha_{JT}$ also reduces, especially in the colder regions near wellbore (see Fig. 4) resulting in lower cooling effects. Furthermore, the presence of impurities in the injected $CO_2$ or contamination of $CO_2$ with the gas already present in the reservoir will also affect the temperature drop and the extent of the cooling zone (Ziabakhsh-Ganji & Kooi, 2014). Gases like $SO_2$ expand the cooling zone while gases like $N_2$ and $CH_4$ tend to contract this zone (Ziabakhsh-Ganji & Kooi, 2014). The Joule-Thomson cooling effect is more pronounced at low-permeability and thin layers, assuming a constant injection rate. With the increase of the reservoir permeability or thickness the impact becomes less and after a certain value the steady-state temperature profile appears to be independent of these parameters.

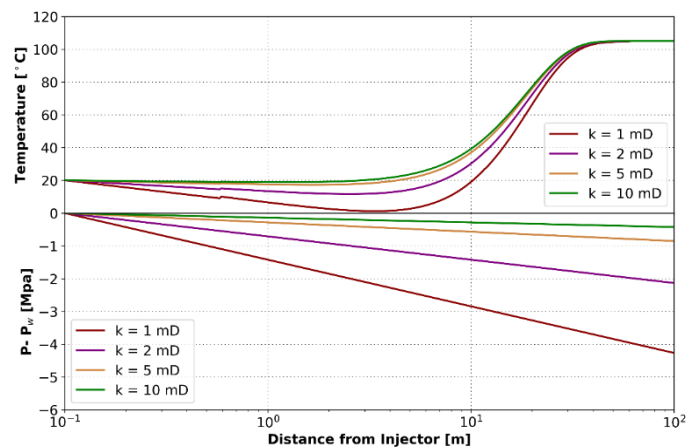

Fig. 13. The impact of permeability on the temperature profiles.

The four values of heat conductivity presented in Fig. 13 correspond to the following values of dimensionless parameter $A^*$: $A^*=2.8\times10^2$, $A^*=1.4\times10^2$, $A^*=5.7\times10^1$, $A^*=2.8\times10^1$, respectively.

6.3 Effect of System Parameters on Minimum Reservoir Temperature

In the sensitivity analysis shown in Fig. 14, only the investigated model parameter was changed while the other parameters remained constant. For simplicity the parameters are normalized to the parameters of the base-case presented in Table 1. Moreover, the $r_{norm}$ on the y-axis is relative to the position of the steady-state temperature of the base case. The normalized distance has a near linear relationship with the mass injection rate, i.e., the higher the injection rate the longer the cold front travels inside the reservoir. In addition, the minimum temperature significantly decreases as the injection rate increases. For example, with doubling of $M_J$, the minimum temperature drops from 10 ºC to -10 ºC, which has huge implications for a $CO_2$ storage project.

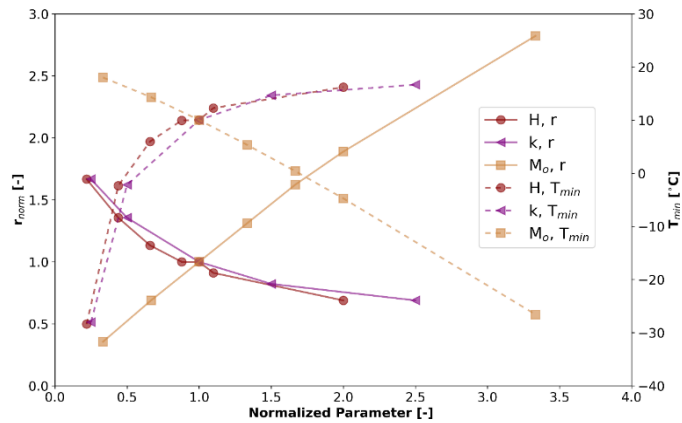

Fig. 14. The impacts of the mass injection, permeability and reservoir thickness on the minimum temperature and the steady-state position of the temperature front in the reservoir. The axes are normalized to the parameters of the base case.

6.4 Sensitivity to Change in the Dimensionless Constants

The dimensionless constants $A^*$ and $V^*$, as defined in Eq. (11) and Eq. (12), are the parameters that account for the J-T effect, and the heat front velocity, respectively. An increase in parameter $A^*$ is equivalent to an increase in the J-T coefficient, injection rate and shale thickness, as well as a decrease in reservoir thickness, shale heat conductivity and permeability. Also, an increase in parameter $V^*$ corresponds to an increase in shale thickness and injection rate, as well as a decrease in shale conductivity.

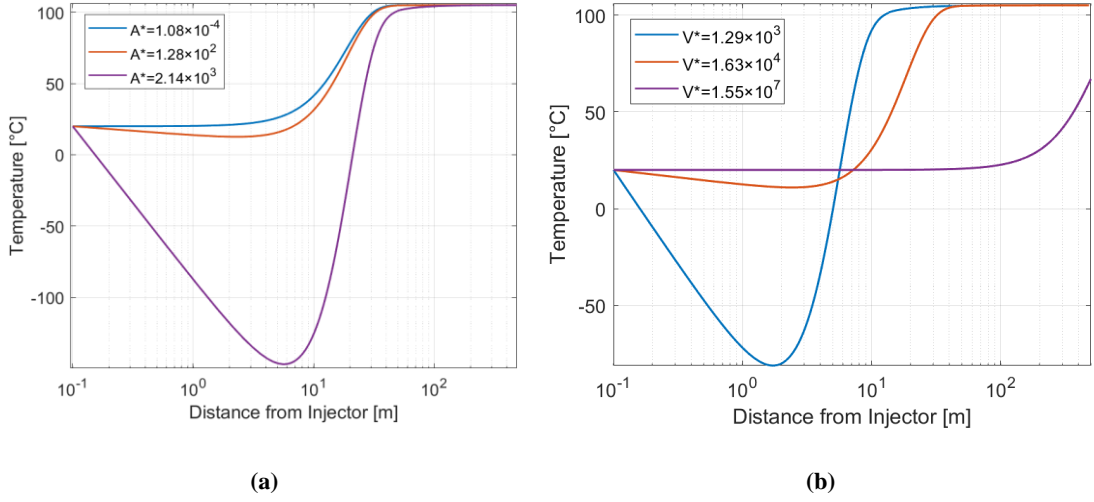

(a)            (b)

Fig. 15. Temperature profile sensitivity to (a) $A^*$ parameter (b) $V^*$ parameter, at time $t_D = 7.1$.

Fig. 15 show the sensitivity of the temperature profile to the dimensionless constants, $A^*$ and $V^*$. From Fig. 15a, it is observed that, while maintaining all other system parameters constant and varying $A^*$, the temperature drop in the system ranges from $\Delta T = 170°C$ to $\Delta T = 0$. In Fig. 15b, varying the values of $V^*$, while holding all other system parameters fixed results to temperature drops varies between $\Delta T = 95°C$ to $\Delta T = 0$. Hence, the system is more sensitive to modifications in parameter $A^*$ when assessing the temperature profiles.

## 7. DISCUSSION

The analytical model presented in this paper (Eq. (19)) assumes a constant injection rate; under given pressure drawdown between the injector and the reservoir. Eq. (17) defines the dimensional parameter for heat front velocity. As temperature front moves significantly slower than the $CO_2$-water front; $V \ll 1$, for which water saturation in the cold region is only slightly above $S_{wi}$. However, varying two-phase mobility during the displacement can affect pressure behaviour, so well injectivity prediction requires accounting for the overall saturation profile. Self-similar solution for displacement of water by $CO_2$ can be matched with the analytical model (Eq. (19)).

Another mechanism requiring two-phase formulation during $CO_2$ injection, which is not accounted for in this study, is mobilisation and migration of natural reservoir fines, that are detached by gas-water menisci (Chequer et al., 2020; Nguyen et al., 2022). The governing system includes the equation for motion of gas-water interface, where fines detachment occurs (Shapiro, 2015). The detachment is determined by the interaction of capillary and DLVO forces exerting the attached fine particle (Yuan & Moghanloo, 2018; Yuan & Moghanloo, 2019). The saturation variation from 1 to $S_{wi}$ around $CO_2$-water front requires generalisation of the overall unsteady-state model for two-phase flow with moving interface (Shapiro, 2016, 2018). Accounting for two-phase flow with fines migration in the model Eq. (19) captures the effects of injectivity decline during $CO_2$ injection.

Other mechanisms contributing to formation damage and injectivity decline, triggered by J-T cooling effect, encompass salt precipitation, rock dissolution, mineral precipitation reaction, and hydrate formation (G Moghanloo et al., 2017; Ge et al., 2020; Nguyen et al., 2021; Turner et al., 2022). Adding those effects into Eq. (19) reflects the effects of formation damage on well injectivity during $CO_2$ storage in aquifers and depleted gas fields. The governing system includes balance equations for energy and mass of salt, methane, and chemical species. In large scale approximation, the hyperbolic system of the quasi-linear equations contains shocks of temperature, saturation, and component concentrations; the corresponding Riemann problems are self-similar (Lake, 1989; Polyanin et al., 2002; Polyanin & Zaitsev, 2003). Accounting for dissipative effects of diffusion and reaction kinetics induces the smoothening of the shocks by travelling wave solutions (Bedrikovetsky, 1993; Polianin & Dil'man, 1994). In several cases, the obtained analytical solutions for water displacement allow for exact and asymptotic upscaling (Cheng & Rabinovich, 2021; Rabinovich et al., 2015); numerous references for upscaling relevant to $CO_2$ injection are available from review (Bedrikovetsky & Borazjani, 2022). In addition, the analytical solution can be obtained for a temperature dependency of the J-T coefficient. This will introduce non-linear term $\alpha_{JT}(T)$ in Eq. (12). For this equation, the characteristics are straight lines. The solution along characteristics is obtained by separation of variables and is given by implicit integral in temperature.

Fig. 9 illustrates how the system parameter $\omega$ impacts the validity of Newton's law within the system. As $\omega$ decreases, the period during which Newton's law remains valid ($t_{D\ max}$) increases, and for the lowest value, there is no such time limit. In general, for every $\varepsilon \ll 1$, there does exist such $\delta$ and $t_\delta$ that for $t < t_\delta$, $T_I - T(z = 0, t) < \varepsilon$. Furthermore, Fig. 10 indicates that for injection processes with specific values for $A^*$, $V^*$, and $\omega$, Newton's law holds over a longer observation period when the radius is larger.

8. CONCLUSIONS

This paper presents an analytical model for a 1D axi-symmetric problem, in which $CO_2$ is injected with constant rate and temperature into a semi-infinite porous medium saturated by gas and water. The model accounts for the Joule-Thomson effect due to "isenthalpic" expansion of the gas and quasi steady-state heat exchange - Newton's law - between the reservoir and the adjacent (over- and under-burden) layers. The exact solution leads to explicit expressions for the temperature profile. The solution shows that the temperature front lags significantly behind the $CO_2$ front due to instant heat exchange between the rock, connate water, and $CO_2$. The exact solution ahead of the front depends on initial temperature and is independent of the injected temperature. In adiabatic reservoir without heat supply from the adjacent layers, the temperature decreases from the well to a minimum value on the heat front, then jumps up on front followed by slow increase up to the initial reservoir temperature. The heat supplied by the adjacent layers has two major impacts on the temperature profile: (1) it decelerates the penetration depth propagation, and (2) it decreases the temperature decline from the well to heat front. Also, the minimum temperature for this case is lower than at the case with no heat gain from the surrounding layers, and its position and value with time remain unchanged. We show that the Newton's heat-exchange law yields tending of the temperature profile to steady-state profile as time tends to infinity; the penetration depth position stabilises. The position of the steady-state temperature strongly depends on dimensionless Joule-Thomson number $A^*$; the dimensionless heat front velocity $V^*$ affects less. The results of this study have major implications for $CO_2$ storage in depleted gas fields and can be used as guide to quantify whether the temperature profile in the reservoir falls within the hydrate formation zone, or whether the induced temperature gradient can jeopardize the mechanical integrity of the rock.

## 9. ACKNOWLEDGEMENTS

The authors thank Shell Global Solutions International for permission to publish this work. Pavel Bedrikovetsky thanks Prof. K. Fedorov (Tyumen University) for fruitful discussions and ARC grant DP190103457 for support.

**Appendix A. Derivation of the exact solution**

Eq. (13) with IC Eq. (14) and BC Eq. (15) present an initial-boundary value problem for a first-order partial differential equation (PDE) which is solved by the method of characteristics (Polyanin & Zaitsev, 2003).

Considering the piston-like displacement of water by $CO_2$, two temperature domains are delineated by a shock front. First, we consider the domain ahead of the heat front, where IC propagates along the characteristic lines. Here the free variable is the dimensionless time, $t_D$.

Assuming that, $T_D = T^*(r_D(t_D), t_D)$, the total derivative of $T^*$ with respect to $t_D$ is obtained as:

$$\frac{dT^*}{dt_D} = \frac{\partial T_D}{\partial r_D}\frac{dr_D}{dt_D} + \frac{\partial T_D}{\partial t_D} = -A^*\frac{1}{r_D^2} - (T^* - 1) \tag{A1}$$

Therefore, the domain ahead of heat front is defined by a system of first order ODE as:

$$\frac{dr_D}{dt_D} = \frac{V^*}{r_D} \; ; \; \frac{dT^*}{dr_D} = \frac{r_D}{V^*}\left(-A^*\frac{1}{r_D^2} - (T^* - 1)\right) \tag{A2}$$

The system of equations Eq. (A2) is solved with the following IC:

$$t_D = 0: T^* = T_D(r_{D0}, 0) = 1 \tag{A3}$$

By integrating the first part of the system of equations Eq. (A2), we obtain the trajectory of characteristics as:

$$t_D = \frac{r_D^2 - 1}{2V^*} - \frac{r_{D0}}{V^*} \tag{A4}$$

Further, we solve the second part of the system of equations Eq (A2) to obtain a general solution for temperature in the domain $t_D < \frac{r_D^2 - 1}{2V^*}$ as:

$$T^* = 1 - \frac{A^*}{2V^*} Ei\left(\frac{r_D^2}{2V^*}\right) e^{-\frac{r_D^2}{2V^*}} + c_1 e^{-\frac{r_D^2}{2V^*}} \tag{A5}$$

Here, the $Ei(x)$ function is defined as:

$$Ei(x) = \int_{-\infty}^{x} \frac{e^t}{t} dt, a < b, \int_{a}^{b} \frac{e^t}{t} dt = \int_{-\infty}^{b} \frac{e^t}{t} dt - \int_{-\infty}^{a} \frac{e^t}{t} dt = Ei(b) - Ei(a) \tag{A6}$$

By using the IC Eq. (A3), the constant of integration $c_1$ in Eq. (A5) is obtained as:

$$c_1 = \frac{A^*}{2V^*} Ei\left(\frac{r_D^2}{2V^*} - t_D\right) \tag{A7}$$

By substituting Eq. (A7) into the general solution Eq. (A5), the temperature profile in the domain ahead of the heat front is obtained as:

$$T_D(r_D, t_D) = 1 - \frac{A^*}{2V^*} Ei\left(\frac{r_D^2}{2V^*}\right) e^{-\frac{r_D^2}{2V^*}} + \left[\frac{A^*}{2V^*} Ei\left(\frac{r_D^2}{2V^*}\right)\right] e^{-\frac{r_D^2}{2V^*}} \tag{A8}$$

Next, we consider the domain behind of the heat front, where BC propagates along the characteristic lines. Here the free variable is the dimensionless radius, $r_D$. Assuming that, $T_D = \bar{T}(r_D, t_D(r_D))$, the total derivative of $\bar{T}$ with respect to $r_D$ is obtained as:

$$\frac{d\bar{T}}{dr_D} = \frac{\partial T_D}{\partial r_D} + \frac{\partial T_D}{\partial t_D}\frac{dt_D}{dr_D} = \frac{1}{V^*}\left(-A^*\frac{1}{r_D^2} - (\bar{T} - 1)\right) \tag{A9}$$

Therefore, the following system of first order ODE is obtained for the domain behind of the heat front.

$$\frac{dt_D}{dr_D} = \frac{r_D}{V^*} \; ; \; \frac{d\bar{T}}{dr_D} = r_D\left[\frac{1}{V^*}\left(-A^*\frac{1}{r_D^2} - (\bar{T} - 1)\right)\right] \tag{A10}$$

In this domain, the system of equations Eq. (A10) is solved with the following BC:

$$r_D = 1: \bar{T} = T_D(0, t_{D0}) = \frac{T_J}{T_I} \tag{A11}$$

By integrating the first part of the system of equations Eq. (A10), the trajectory of characteristics is:

$$t_D = \frac{r_D^2 - 1}{2V^*} + t_{D0} \tag{A12}$$

Additionally, we solve the second part of the system of equations Eq (A10) to obtain a general solution for temperature in the domain $t_D > \frac{r_D^2 - 1}{2V^*}$ as:

$$\bar{T} = 1 - \frac{A^*}{2V^*}Ei\left(\frac{r_D^2}{2V^*}\right)e^{-\frac{r_D^2}{2V^*}} + c_2 e^{-\frac{r_D^2}{2V^*}} \tag{A13}$$

By using the BC Eq. (A11), the constant of integration $c_2$ in Eq. (A13) is obtained as:

$$c_2 = \left(\frac{T_J - T_I}{T_I}\right)e^{\frac{1}{2V^*}} + \frac{A^*}{2V^*}Ei\left(\frac{1}{2V^*}\right) \tag{A14}$$

By substituting Eq. (A14) into the general solution Eq. (A13), the temperature profile in the domain behind the heat front is obtained as:

$$T_D(r_D, t_D) = 1 - \frac{A^*}{2V^*}Ei\left(\frac{r_D^2}{2V^*}\right)e^{-\frac{r_D^2}{2V^*}} + \left[\left(\frac{T_J - T_I}{T_I}\right)e^{\frac{1}{2V^*}} + \frac{A^*}{2V^*}Ei\left(\frac{1}{2V^*}\right)\right]e^{-\frac{r_D^2}{2V^*}} \tag{A15}$$

Finally, the shock emanates from the point (1,0) in the $r_D$ - $t_D$ plane; which allows for the overall dimensionless solution for the problem defined by Eq. (13) with IC Eq. (14) and BC Eq. (15) to be obtained as:

$$T_D(r_D, t_D) = \begin{cases} 1 - \frac{A^*}{2V^*} Ei\left(\frac{r_D^2}{2V^*}\right) e^{-\frac{r_D^2}{2V^*}} + \left[\left(\frac{T_J - T_I}{T_I}\right) e^{\frac{1}{2V^*}} + \frac{A^*}{2V^*} Ei\left(\frac{1}{2V^*}\right)\right] e^{-\frac{r_D^2}{2V^*}}; & t_D > \frac{r_D^2 - 1}{2V^*} \\ 1 - \frac{A^*}{2V^*} Ei\left(\frac{r_D^2}{2V^*}\right) e^{-\frac{r_D^2}{2V^*}} + \left[\frac{A^*}{2V^*} Ei\left(\frac{r_D^2}{2V^*} - t_D\right)\right] e^{-\frac{r_D^2}{2V^*}}; & t_D < \frac{r_D^2 - 1}{2V^*} \end{cases} \quad (A16)$$

The substitution of the dimensionless parameters in Eqs. (10-12) into Eq. (A16) yields the dimensional form of the solution as:

$$T(r,t) = \begin{cases} T_I + (T_J - T_I) e^{-\frac{B}{V}\pi(r^2 - r_w^2)} + \frac{\alpha_{JT}\mu_f q_J}{4\pi k_{rf}^e kH} \left[Ei\left(\frac{B}{V}\pi r_w^2\right) - Ei\left(\frac{B}{V}\pi r^2\right)\right] e^{-\frac{B}{V}\pi r^2}; & \pi r^2 < \pi r_w^2 + V q_J t \\ T_I + \frac{\alpha_{JT}\mu_f q_J}{4\pi k_{rf}^e kH} \left\{Ei\left(\frac{B}{V}(\pi r^2)\right) - Ei\left(\frac{B}{V}\pi r^2\right)\right\} e^{-\frac{B}{V}\pi r^2}; & \pi r^2 > \pi r_w^2 + V q_J t \end{cases}$$

(A17)

**Appendix B. Exact solution for heat exchange in the adjacent layers of semi-infinite thickness**

The assumptions of the quasi 2D heat conductivity problem in adjacent layers comprise: No horizontal heat transfer, i.e. $\gamma_x = 0$; constant vertical heat conductivity $\gamma_z = \gamma = const$; thickness of the shale is much smaller than the thickness of the reservoir and the adjacent layers. The equation that governs heat transfer along the vertical direction in the adjacent formations above and below the reservoir is expressed in the form:

$$(\phi_a \rho_w c_w + (1 - \phi_a)\rho_{ra} c_{ra}) \frac{\partial T(r,z,t)}{\partial t} = (\phi_a \gamma_w + (1 - \phi_a)\gamma_{ra}) \frac{\partial^2 T(r,z,t)}{\partial z^2}, \quad 0 < z < \infty \quad (B1)$$

Here, $\phi_a$ is adjacent layers porosity, $\rho_{ra}$ [ML$^{-3}$] is adjacent layers rock density, $c_{ra}$ [L$^2$T$^{-2}$K$^{-1}$] is the specific heat capacity of adjacent layers rock, $\gamma_w$ [MLT$^{-3}$K$^{-1}$] is thermal conductivity of water, $\gamma_{ra}$ [MLT$^{-3}$K$^{-1}$] is thermal conductivity of adjacent layers and $z$ [L] is distance in the vertical direction.

Further, the thermal diffusivity during the heat transfer is defined as:

$$a_a = \frac{\phi_a \gamma_w + (1 - \phi_a)\gamma_{ra}}{\phi_a \rho_w c_w + (1 - \phi_a)\rho_{ra} c_{ra}} = \frac{\gamma^a}{c^a} \quad (B2)$$

Substituting Eq. (B2) into Eq. (B1), the heat transfer equation is reformulated to the form:

$$\frac{\partial T(r,z,t)}{\partial t} = a_a \frac{\partial^2 T(r,z,t)}{\partial z^2}, \quad 0 < z < \infty \quad (B3)$$

The initial condition (IC) corresponding to the equality of temperature in all formations is given as:

$$T(r, z, t = 0) = T_I \quad (B4)$$

Additionally, the boundary conditions (BCs) for the semi-infinite vertical domain is defined as:

$$T(r, z \to \infty, t) = T_I \tag{B5}$$

$$\frac{\partial T}{\partial z}(r, z = 0, t) = -\frac{\gamma^s}{\gamma^a}\frac{1}{l}[T(r = const., t) - T(r, z = 0, t)] \tag{B6}$$

Here, $\gamma^s$ denotes the overall shale heat conductivity and it is given as:

$$\gamma^s = \phi_s \gamma_w + (1 - \phi_s)\gamma_{rs} \tag{B7}$$

Also, $T(r = const., t)$ denotes the temperature in the reservoir post injection (taken at constant radial distances from injection well). We further define the overall reservoir volumetric heat capacity as:

$$c^{res} = \phi(1 - S_{wi})\rho_f c_f + \phi S_{wi} \rho_w c_w + (1 - \phi)\rho_s c_s \tag{B8}$$

In this work, we introduce a new dimensionless parameter, $\omega = f(\gamma, c, H, l)$, which establishes a ratio of for the following properties; heat conductivity, heat capacity and thickness of the reservoir and the shale

$$\omega = \sqrt{\frac{c^{res}\gamma^s H}{c^a \gamma^a l}} \tag{B9}$$

In addition, we consider the following dimensionless variables to nondimensionalize Eqs. (B3 - B6):

$$t_D = \frac{\gamma^s t}{Hlc^{res}} \; ; \; z_D = \frac{z}{\sqrt{\frac{Hlc^{res}\gamma^a}{\gamma^s \; c^a}}} \; ; \; \theta = \frac{T}{T_I} \tag{B10}$$

Using Eq. (B10), the second order PDE Eq (B3) together with the IC Eq. (B4) and BCs Eqs. (B5 - B6) are respectively expressed in dimensionless form as:

$$\frac{\partial \theta}{\partial t_D} = \frac{\partial^2 \theta}{\partial z_D^2}, \; 0 < z_D < \infty \tag{B11}$$

$$\theta(z_D, t_D = 0) = 1 \tag{B12}$$

$$\theta(z_D \to \infty, t_D) = 1 \tag{B13}$$

$$\frac{\partial \theta}{\partial z_D}(z_D = 0) - \omega\theta(z_D = 0) = -\omega T_D(r_D = const.) \tag{B14}$$

The details of the solution to the problem defined by Eqs. (B11 - B14) is expounded in the work of Polyanin and Nazaikinskii (2016). Based on the approach discussed by Polyanin and Nazaikinskii (2016), the general solution to the second order problems of this kind is obtained as:

$$T(z,t) = \int_0^\infty f(x) G(z,x,t) dx - a_a \int_0^t g(\tau) G(z,0,t-\tau) d\tau \tag{B15}$$

Here, $f(x)$ and $g(\tau)$ are arbitrary functions of $x$ and $\tau$, respectively and the function $G(z,x,t)$ is defined as:

$$G(z,x,t) = \frac{1}{2\sqrt{\pi a_a t}}\left\{ e^{-\frac{(z-x)^2}{4a_a t}} + e^{-\frac{(z+x)^2}{4a_a t}} - 2k\left[\sqrt{\pi a_a t}\, e^{a_a k^2 t + k(z+x)} erfc\left(\frac{z+x}{2\sqrt{a_a t}} + k\sqrt{a_a t}\right)\right]\right\} \tag{B16}$$

$$G(z,0,t-\tau) = \frac{1}{\sqrt{\pi a_a(t-\tau)}}\left\{ e^{-\frac{z^2}{4a_a(t-\tau)}} - k\left[\sqrt{\pi a_a(t-\tau)}\, e^{a_a k^2 (t-\tau) + kz} erfc\left(\frac{z}{2\sqrt{a_a(t-\tau)}} + k\sqrt{a_a(t-\tau)}\right)\right]\right\} \tag{B17}$$

From the general solution given by Eq. (B15), together with Eqs. (B16 - B17), the solution to the heat transfer problem Eqs. (B11 - B14) is obtained as:

$$\theta(z_D, t_D) = erf\left(\frac{z_D}{2\sqrt{t_D}}\right) + e^{\omega^2 t_D + \omega z_D} erfc\left(\frac{z_D}{2\sqrt{t_D}} + \omega\sqrt{t_D}\right) +$$

$$\omega \int_0^{t_D} \frac{T_D(r_D = const., \tau)}{\sqrt{\pi(t_D - \tau)}}\left\{ e^{-\frac{z_D^2}{4(t_D - \tau)}} - \omega\left[\sqrt{\pi(t_D - \tau)}\, e^{\omega^2(t_D - \tau) + \omega z_D} erfc\left(\frac{z_D}{2\sqrt{(t_D - \tau)}} + \omega\sqrt{(t_D - \tau)}\right)\right]\right\} d\tau \tag{B18}$$

The value for temperature $\theta(z_D, t_D)$ on the interface between an adjacent layers and shale is obtained by setting $z_D = 0$ in Eq. (B18) to obtain:

$$\theta(z_D = 0, t_D) = e^{\omega^2 t_D} erfc(\omega\sqrt{t_D}) +$$

$$\omega \int_0^{t_D} \frac{T_D(r_D = const., \tau)}{\sqrt{\pi(t_D - \tau)}}\left\{1 - \omega\left[\sqrt{\pi(t_D - \tau)}\, e^{\omega^2(t_D - \tau)} erfc(\omega\sqrt{(t_D - \tau)})\right]\right\} d\tau \tag{B19}$$

where,

$$T_D(r_D = const., \tau) = \begin{cases} 1 + \left[\left(\frac{T_J - T_I}{T_I}\right) e^{\frac{1}{2V^*}} + \frac{A^*}{2V^*}\left(Ei\left(\frac{1}{2V^*}\right) - Ei\left(\frac{r_D^2}{2V^*}\right)\right)\right] e^{-\frac{r_D^2}{2V^*}} & t_D > \frac{r_D^2 - 1}{2V^*} \\ 1 - \frac{A^*}{2V^*} e^{-\frac{r_D^2}{2V^*}}\left[Ei\left(\frac{r_D^2}{2V^*}\right) - Ei\left(\frac{r_D^2}{2V^*} - \tau\right)\right] & \frac{r_D^2 - 1}{2V^*} < t_D \end{cases} \tag{B20}$$